\documentstyle[12pt]{article}
\begin{document}
{\pagestyle{empty}
%\vskip 1.5cm
\vskip 2.5cm
~~\\

{\renewcommand{\thefootnote}{\fnsymbol{footnote}}
\centerline{\large \bf Protein Folding Simulations in a Deformed Energy 
                    Landscape}
}
\vskip 3.0cm

\centerline{Ulrich H.E.~Hansmann
\footnote{\ \ e-mail: hansmann@mtu.edu}}
\vskip 1.5cm
\centerline{ {\it Department of Physics}}
\centerline{{\it Michigan Technological University}}
\centerline{{\it Houghton, MI 49931-1295}}

\medbreak
\vskip 3.5cm
 
\centerline{\bf ABSTRACT}
\vskip 0.3cm
A modified version of stochastic tunneling, a recently
introduced global optimization technique, is introduced as a
new generalized-ensemble technique and tested for a benchmark
peptide, Met-enkephalin. It is demonstrated that the new technique allows 
to evaluate  folding properties and especially the glass temperature $T_g$ 
of this peptide. 

\vskip 3.5cm

\noindent
{\it Key words:} Generalized-Ensemble Simulations, Protein Folding,
                 Stochastic Tunneling.
 
\vfill
\newpage}
\baselineskip=0.8cm
%\noindent
Numerical simulations of  biological molecules can be 
 extremely difficult when the molecule is described by ``realistic''
energy functions where  interactions between all atoms are taken into
account. For a large class of molecules, in particular for 
peptides or proteins,  
the various competing interactions lead to frustration and a rough
energy landscape. At low temperatures canonical simulations will
get trapped in one of the  multitude of local minima separated
by high energy barriers  and 
 physical quantities cannot be calculated accurately. 
One way  to overcome this difficulty in protein simulations 
is by utilizing   so-called
{\it generalized ensembles}\cite{MyReview}, which are based on a non-Boltzmann
probability distribution.
Multicanonical sampling \cite{MU} and simulated tempering \cite{L} 
are prominent examples of such an approach. 
Application of  these techniques to the protein folding  problem was first 
addressed in Ref.~\cite{HO} and their usefulness 
 for simulation of biological molecules and other
complex systems \cite{HO}-\cite{hoo98b} has become increasingly recognized. 

However, generalized-ensemble methods are not without problems.
 In contrast to  canonical  simulations
the weight factors are not {\it a priori} known. 
Hence, for a computer experiment one needs estimators of the  weights, and
the problem of finding good estimators  is  often limiting the use
of generalized-ensemble techniques. Here we describe and test a new 
generalized ensemble where determination of the weights is by construction 
of the ensemble simple and straightforward. Our method is based on a recently
introduced global optimization technique, stochastic tunneling \cite{HW}.

Canonical simulations of proteins at low temperature  are hampered by 
the roughness of the potential energy surface: local minima
 are separated by high energy barriers. To enhance
sampling we propose  to weight conformations  not with 
the Boltzmann factor $w_B(E) = \exp(-E/k_BT)$, but with a weight
\begin{equation}
w_f(E) = \exp(f(E)/k_BT)~.
\label{eqwe}
\end{equation}
Here, $T$ is a low temperature, $k_B$ the Boltzmann constant, and 
 $f(E)$ is a non-linear transformation of the potential energy onto the
interval $[0,-1]$ chosen such that the relative 
location of all minima is preserved. 
The physical idea  behind such an approach is to allow the system 
at a given low temperature $T$  to ``{\it tunnel}'' 
through energy barriers of arbitrary height, while the low energy region is 
still well resolved. A transformation with the above characteristics 
can be realized by
\begin{equation}
 \quad f_1(E)=-e^{-(E-E_0)/n_F}~.
\label{f1}
\end{equation}
Here,  $E_0$ is  an estimate for the ground state and $n_F$ the number of 
degrees of freedom of the system.  Eq.~\ref{f1} is a special choice of 
the transformation recently introduced under the name 
``stochastic tunneling'' \cite{HW} to the corresponding problem 
of global minimization in complex potential energy landscapes. 
One can easily find further examples for transformations with the above stated
properties, for instance, 
\begin{equation}
 \quad f_2(E)=-(1+ (E-E_0)/n_F)^{-1}~.
\label{f2}
\end{equation}
We will restrict our investigation to these two transformations without 
claiming that they are an optimal choice.

A simulation in the 
above ensemble, defined by the weight of Eq.~\ref{eqwe} with a suitable
chosen non-linear transformation $f(E)$, 
will sample a broad range of energies. Hence, application of re-weighting
technique \cite{FS} allows to calculate the expectation value of any physical
quantity ${\cal O}$ over a large range of temperatures $T$ by
\begin{equation}
< {\cal O} >_T ~= \frac{
                   \displaystyle{\int dE~ {\cal O} (E) P_f(E) 
                                          w^{-1}_f (E) e^{-E/k_BT}}  }
                  {\displaystyle{\int dE~ P_f(E)w^{-1}_f (E) e^{-E/k_BT}}}~.
\label{reweight}
\end{equation}
In this point our method is similar to other generalized-ensemble
techniques such as the multicanonical sampling \cite{MU}, however,
our method differs from them in that
 the weights are explicitly given by Eq.~\ref{eqwe}. 
One only needs to find  an estimator for the ground-state energy $E_0$
in the transforming functions $f_1(E)$ or $f_2(E)$ 
(see  Eqs.~\ref{f1} and \ref{f2}) 
which in earlier work \cite{H97a,HO96g} was found to be 
much easier than the determination of weights for 
 multicanonical algorithm \cite{MU} 
or simulated tempering \cite{L}.

The  new simulation technique was tested for 
 Met-enkephalin, one of the simplest peptides, which has become  a often
used model to examine new algorithms. 
Met-enkephalin has the amino-acid sequence Tyr-Gly-Gly-Phe-Met.
The potential energy function
$E_{tot}$ that was used is given by the sum of
the electrostatic term $E_{es}$, 12-6 Lennard-Jones term $E_{vdW}$, and
hydrogen-bond term $E_{hb}$ for all pairs of atoms in the peptide together with
the torsion term $E_{tors}$ for all torsion angles:
\begin{eqnarray}
E_{tot} & = & E_{es} + E_{vdW} + E_{hb} + E_{tors},\\
E_{es}  & = & \sum_{(i,j)} \frac{332q_i q_j}{\epsilon r_{ij}},\\
E_{vdW} & = & \sum_{(i,j)} \left( \frac{A_{ij}}{r^{12}_{ij}}
                                - \frac{B_{ij}}{r^6_{ij}} \right),\\
E_{hb}  & = & \sum_{(i,j)} \left( \frac{C_{ij}}{r^{12}_{ij}}
                                - \frac{D_{ij}}{r^{10}_{ij}} \right),\\
E_{tors}& = & \sum_l U_l \left( 1 \pm \cos (n_l \chi_l ) \right),
\end{eqnarray}
where $r_{ij}$ is the distance between the atoms $i$ and $j$,
and $\chi_l$ is the $l$-th torsion angle.
The parameters ($q_i,A_{ij},B_{ij},C_{ij},
D_{ij},U_l$ and $n_l$) for the energy function were adopted
from ECEPP/2.\cite{EC}

The computer code SMC\footnote{The program SMC was written by
Dr.~Frank Eisenmenger (eisenmenger@rz.hu-berlin.de)}  was used. 
The simulations were started from  completely random initial
conformations (Hot Start) and  one Monte Carlo sweep updates every torsion angle
of the peptide once. The peptide bond angles $\omega$ were fixed 
 to their common value $180^{\circ}$, which
 left  19 torsion angles ($\phi,~\psi$, and $\chi$) as independent
degrees of freedom (i.e., $n_F = 19$). The interaction
of the peptide with the solvent was neglected in the simulations and 
 the dielectric constant $\epsilon$ set equal to 2. 
In short preliminary runs it was found that $T=8$ K was the optimal temperatures
for simulations relying on the transformation $f_1(E)$ (Eq.~\ref{f1}), and 
$T=6 K$ for simulations relying on the
second chosen transformation $f_2(E)$ (Eq.~\ref{f2}).  The free parameter
$E_0$ was set in Eq.~\ref{f1} or (\ref{f2}) to $E_0= -10.72$
kcal/mol, the ground state energy as known from previous work. 
In addition, simulations were also performed where $E_0$ was  dynamically
updated in the course of the simulation  and  set to the lowest ever
encountered energy. In these runs the (known) ground state was
found in less than 5000 MC sweeps. Hence,  determination
of the weights is easier than in other generalized-ensemble techniques since
 in earlier work\cite{HO} it was found that at  least 40,000 sweeps
were needed to calculate multicanonical weights. We remark that a
Monte Carlo sweep in both algorithm takes approximately the
same amount of CPU time.

All thermodynamic quantities were then  calculated from
a single production run of 1,000,000 MC sweeps which followed 10,000
sweeps for thermalization. At the end of every  sweep we stored
the energies of the conformation and 
the radius of gyration 
\begin{equation}
   R = \frac{1}{N_{atoms}^2} \sum_{i,j}^{N_{atoms}} (\vec{r}_i - \vec{r}_j)^2
\label{Rgyr}
\end{equation}
for further analyses.

In order to demonstrate the dynamical behavior 
of the algorithm the ``time series'' and histograms of potential energy 
are shown for both choices of the
transforming functions $f_1(E)$ (Fig.~1) and $f_2(E)$ (Fig.~2). 
Both choices of the
non-linear transformation with which the energy landscape was deformed 
in the simulations lead to qualitatively the same picture. 
In Fig.~1a and Fig.~2a, respectively, one can  see that  the whole energy range
between $E < -10$ kcal/mol (the ground state region) and $E\approx 20$ kcal/mol
(high-energy, coil states) is sampled. However, unlike in the multicanonical
algorithm the energies are not sampled uniformly and low-energy states
appear with higher frequency than high energy states. 
 However, as one can see from the logarithmic scale of 
 Fig.~1b and 2b where the histograms are displayed for these simulations, 
 high-energy states are only suppressed by
three orders of magnitude and their probability is still
   large enough  to allow crossing of energy barriers.  
 Hence large parts of the configuration space 
are sampled by our method and it is justified to 
calculate from these simulations  thermodynamic quantities  by means of
 re-weighting, see Eq.~\ref{reweight}.

Here, the average radius of  gyration $<R>$, 
which is  is a measure for the compactness of protein configurations and
defined in Eq.~\ref{Rgyr}, was calculate for various temperatures. 
 In Fig.~3 the results for  the new ensemble,
using the defining non-linear transformations $f_1(E)$ or $f_2(E)$, 
are compared with
the ones of a multicanonical run with equal number of Monte Carlo sweeps. 
As one can see,
the values of $<R>(T)$ agree for all three simulations over the whole
temperature range. Hence, it is obvious that simulations in the new ensemble
are indeed  well able 
to  calculate thermodynamic averages over a wide temperature
range. 

After having established the new techniques as 
 a possible alternative to  other 
generalized-ensemble techniques such as multicanonical 
sampling or simulated tempering,  its usefulness shall be further demonstrated 
 by calculating   the free energy of Met-enkephalin 
as a function of  $R$:
\begin{equation}
  G(R) = -k_B T \log P (R)
\end{equation}
where 
\begin{equation}
  P(R) = P_{f}(R) * w_{f}^{-1} (E(R)) e^{-E(R)/k_BT}~.
\end{equation}
Here,  a normalization is chosen where  the minimal value of
$G_{min}(R) = 0$.  The chosen temperature  was $T=230$K, which
was found in earlier work \cite{HMO97b} as the folding temperature
$T_f$ of Met-enkephalin. The results,  which rely on 
the transformation $f_1(E)$ of the energy landscape given by Eq.~\ref{f1}
are displayed in Fig.~4. At this temperature one observes clearly a ``funnel'' 
towards low values of $R$ which correspond to compact structures. Such
a funnel-like landscape was already observed in Ref.~\cite{hoo98b} 
for Met-enkephalin, utilizing a different set of order parameters,  and
is  predicted by the landscape theory of folding \cite{DC}.

The essence of the funnel landscape idea is competition between the
tendency towards the folded state and trapping due to ruggedness of
the landscape. One way to measure this  competition is by the
ratio \cite{tang}:
\begin{equation}
Q = \frac{\overline{E-E_0}}{\sqrt{\overline{E^2} - \bar{E}^2}}~,
\label{eq_Q}
\end{equation}
where the bar denotes averaging over compact configurations.
The landscape theory asserts that good folding protein sequences 
are characterized by  large values of $Q$ \cite{tang}. Using the results
of our simulations and defining a compact structure as one where
$R(i) \le  23 \AA$, we find $\overline{E-E_0} = 13.96(3)$ Kcal/mol,
$\overline{E^2} - \bar{E}^2 = 0.49(2)$, from which we estimate for
 the above ratio 
$Q=20.0(5)$. This value indicates that Met-enkephalin is good folder and
is consistent with earlier work \cite{HMO97b} where we evaluated an
alternative characterization of 
folding properties. Thirumalai and collaborators \cite{KTh}
have conjectured
that  the kinetic accessibility of the native conformation can be classified
by the parameter
\begin{equation}
\sigma = \frac{T_{\theta} - T_f}{T_{\theta}}~,
\label{sig}
\end{equation}
i.e., the smaller $\sigma$ is, the more easily a protein can fold.
Here $T_f$ is the folding temperature and $T_{\theta}$ the collapse
temperature. With  values for $T_{\theta} = 295$ K  and $T_f = 230$ K,
as measured in Ref.~\cite{HMO97b},  
one has for Met-enkephalin $\sigma \approx 0.2$, indicating again 
 that the peptide has  good folding properties.

Yet another characterization of folding properties relies on knowledge
of the glass temperature $T_g$ and is closely related to  Eq.~\ref{eq_Q}.
As the number of available states  gets reduced with the decrease
of temperature, the possibility of local trapping increases
substantially. Glassy
behavior appears when the residence time in some local traps
becomes of the order of the folding event. Folding dynamics is now
non-exponential since different traps have different escape
times~\cite{REF1}. For temperatures above the glass
transition temperature $T_g$, the folding dynamics is exponential and
a configurational diffusion coefficient average the effects of the
short lived traps \cite{SOW}. It is expected
that for a good folder the glass transition 
temperature, $T_g$, where glass behavior
sets in, has to be significantly lower than the folding temperature
$T_f$, i.e. a good folder can be characterized by the
relation \cite{BOSW}
\begin{equation}
\frac{T_f}{T_g} > 1~.
\end{equation}
I present here for the first time a numerical estimate of this glass
transition temperature for the peptide Met-enkephalin. The calculation of
the estimate is based on the approximation \cite{BOSW}
\begin{equation}
T_g = \sqrt{\displaystyle{\frac{\overline{E^2} -\bar{E}^2}{2 k_B S_0}}}~,
\end{equation}
where the bar indicates again averaging over compact structures
and $S_0$ is the entropy of these states estimated by the relation
\begin{equation}
 S_0 = \frac{\overline{\log w(i)}}{\overline{w(i)}} - \log \tilde{z} -C
\end{equation} 
Here, $\tilde z = \sum_{compact} w(i)$ and $C$ chosen such that 
the entropy of the ground state becomes zero. The results of  the
simulation in the new ensemble defined by the transformation $f_1(E)$,
leads to a value of $s_0 =2.3(7)$. Together with the above quoted value 
for $\overline{E^2} -\bar{E}^2 = 0.49(2)$ (in (Kcal/mol)$^2$) one
therefore finds as an estimate for the glass transition temperature
\begin{equation}
 T_g = 160 (30)~{\rm K}~.
\end{equation}
Since it was found in earlier work \cite{HMO97b} that $T_f=230(30)$,
it is obvious that the ratio $T_f/T_g > 1$ and again one
finds that Met-enkephalin has good folding properties. Hence, we
see that  there is a strong correlation between all three folding criteria.

Let me  summarize my results. I have proposed to utilize a recently
introduced global optimization technique, stochastic tunneling, in 
such a way that it allows calculation of thermodynamic quantities.
The new generalized-ensemble technique was tested for a benchmark
peptide, Met-enkephalin. It was demonstrated that the new technique allows 
to evaluate the folding properties of this peptide and 
 an estimate for the glass transition temperature $T_g$
in that system was presented. Currently I am evaluating the efficiency of 
the new method in simulations of larger molecules.

\vspace{0.5cm}
\noindent
{\bf Acknowledgements}: \\
This article was written in part when I was visitor at the  Institute of
Physics, Academia Sinica, Taipei, Taiwan. I like to thank the Institute
and specially C.K. Hu, head of the Laboratory for Statistical and
Computational Physics, for the kind hospitality extended to me.
Financial support from  a Research Excellence
Fund  of the State of Michigan is  gratefully acknowledged.\\

%%%%%%%%%%%%%%%%%%%%%%%%% references %%%%%%%%%%%%%%%%%%%

\newpage
\noindent
{\bf \Large FIGURE CAPTIONS:}\\
\begin{enumerate}
\item ``Time series''(a) of potential energy $E$ of Met-enkephalin
      for a simulation in a  generalized ensemble
      defined by the transformation $f_1(E)$ of Eq.~\ref{f1}  and
      the corresponding histogram (b) of potential energy.
\item ``Time series''(a) of potential energy $E$ of Met-enkephalin
      for a simulation in a  generalized ensemble
      defined by the transformation $f_2(E)$ of Eq.~\ref{f2} (a)  and 
      the corresponding histogram (b) of potential energy.
\item Average radius of gyration $<R>$ (in $\AA^2$) as a function of 
      temperature (in $K$). The results of a multicanonical simulation 
      of 1,000,000 MC sweeps were compare with simulations of equal
      statistics in the new ensemble utilizing
      either the no-linear transformation $f_1(E)$ or $f_2(E)$. 
\item Free energy $G(R)$ as a function of the radius of gyration $R$
      for   $T=230$ K. The results rely on a
      generalized-ensemble simulation based on the transformation $f_1(E)$ of
      the energy landscape  s defined in Eq.~\ref{f1}.
\end{enumerate}

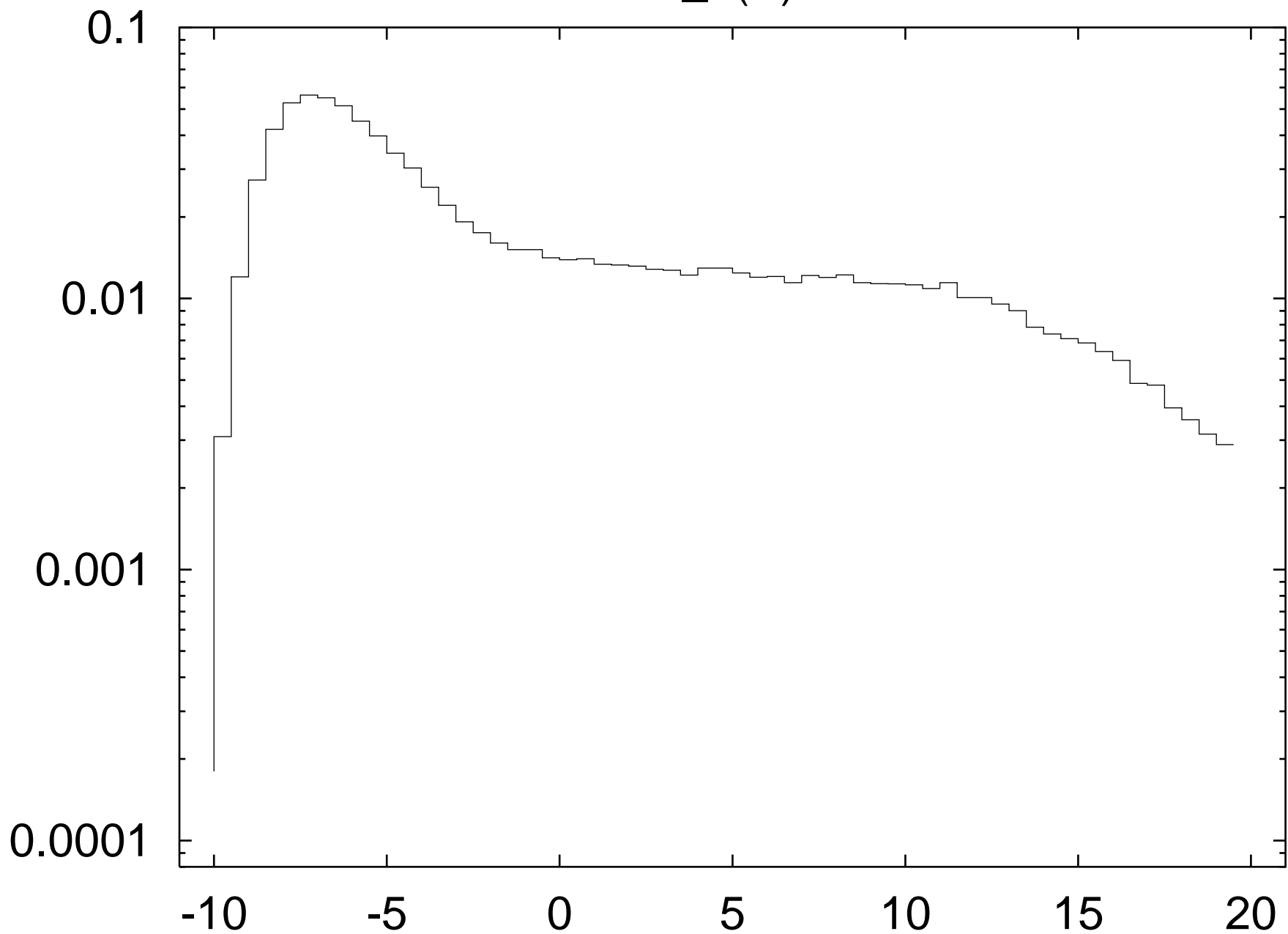
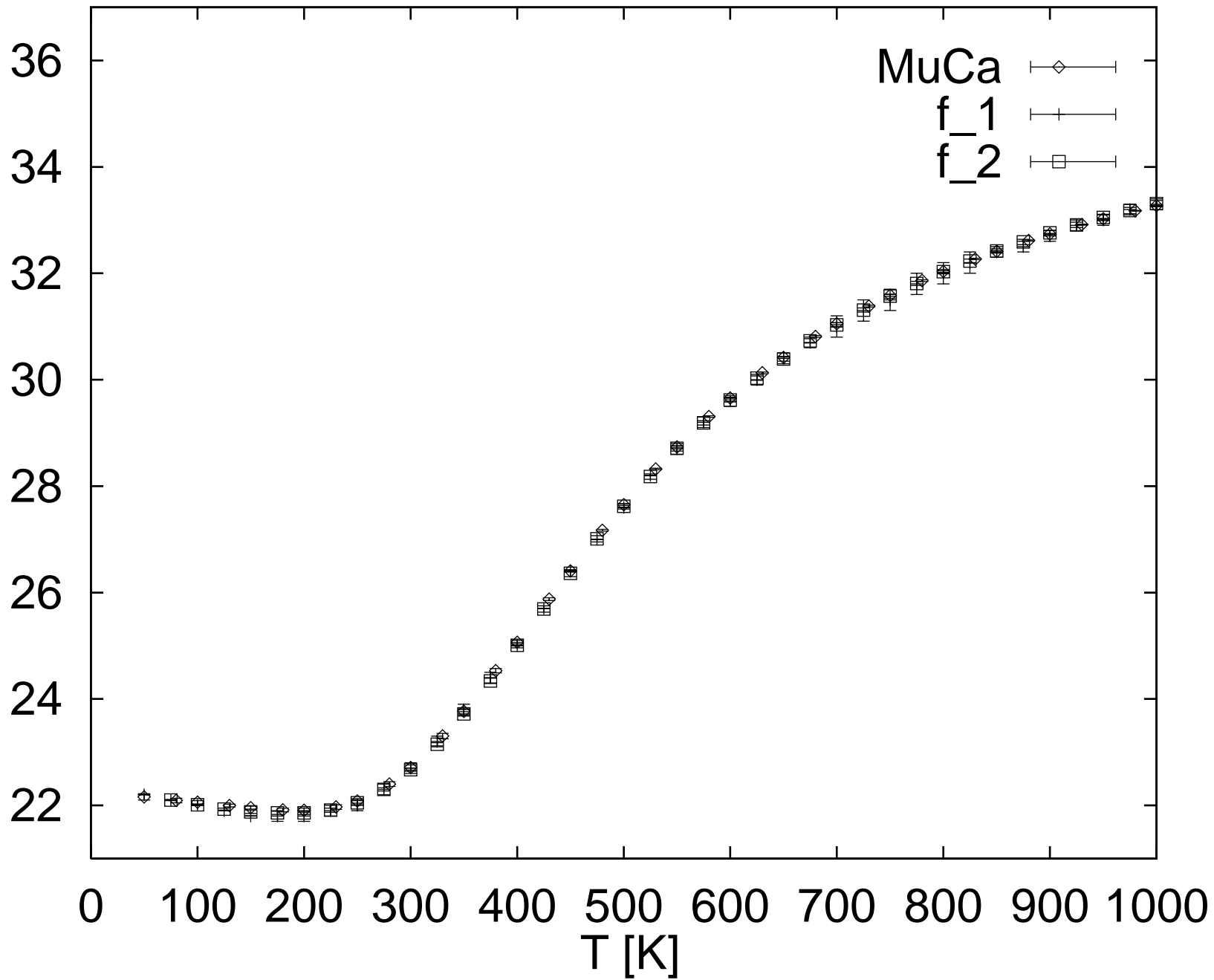
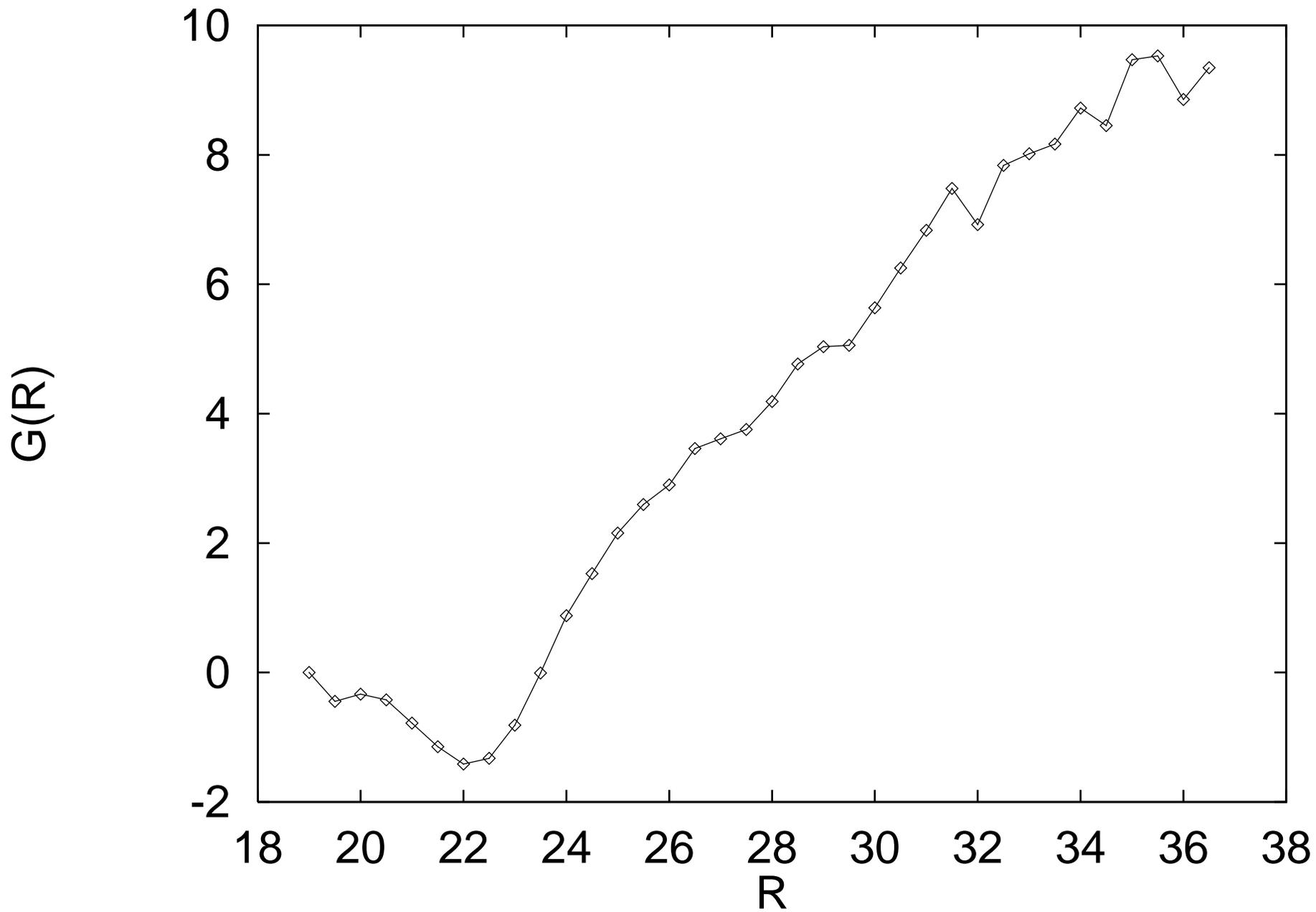
\end{document}